# Diabatic gates for frequency-tunable superconducting qubits

R. Barends,[1, *] C. M. Quintana,[1, *] A. G. Petukhov,[2, *] Yu Chen,[1] D. Kafri,[2] K. Kechedzhi,[2] R. Collins,[1] O. Naaman,[1] S. Boixo,[2] F. Arute,[1] K. Arya,[1] D. Buell,[1] B. Burkett,[1] Z. Chen,[1] B. Chiaro,[3] A. Dunsworth,[1] B. Foxen,[3] A. Fowler,[1] C. Gidney,[1] M. Giustina,[1] R. Graff,[1] T. Huang,[1] E. Jeffrey,[1] J. Kelly,[1] P. V. Klimov,[1] F. Kostritsa,[1] D. Landhuis,[1] E. Lucero,[1] M. McEwen,[3] A. Megrant,[1] X. Mi,[1] J. Mutus,[1] M. Neeley,[1] C. Neill,[1] E. Ostby,[2] P. Roushan,[1] D. Sank,[1] K. J. Satzinger,[1] A. Vainsencher,[1] T. White,[1] J. Yao,[1] P. Yeh,[1] A. Zalcman,[1] H. Neven,[2] V. N. Smelyanskiy,[2] and John M. Martinis[1, 3]

[1]*Google, Santa Barbara, CA 93117, USA*
[2]*Google, Venice, CA 90291, USA*
[3]*Department of Physics, University of California, Santa Barbara, CA 93106, USA*
(Dated: July 4, 2019)

We demonstrate diabatic two-qubit gates with Pauli error rates down to $4.3(2) \cdot 10^{-3}$ in as fast as 18 ns using frequency-tunable superconducting qubits. This is achieved by synchronizing the entangling parameters with minima in the leakage channel. The synchronization shows a landscape in gate parameter space that agrees with model predictions and facilitates robust tune-up. We test both iSWAP-like and CPHASE gates with cross-entropy benchmarking. The presented approach can be extended to multibody operations as well.

One of the key goals in quantum computing is developing gates that are fast and precise. For superconducting qubits, recent advances have led to gate speeds that are up to three orders of magnitude faster than typical coherence times, enabling pilot experiments in simulation and error correction [1–6]. However, while fast gates minimize error from decoherence, they are prone to imprecision in quantum control and can cause leakage out of the computational subspace. Superconducting qubits are by nature multi-level systems whose higher level states can become populated. For weakly nonlinear qubits such as the transmon [7], this leakage is found to be long-lived [8] and can grow to significant levels when running long algorithms, necessitating mitigation techniques for quantum error correction [9]. The need to minimize leakage has led to the development of adiabatic gates [10] as well as the introduction of resonators to limit interactions between qubits [11, 12], both of which increase gate duration and decoherence error. The demonstration of a fast gate that balances low control error and low leakage remains an open challenge.

Here, we report diabatic gates that have minimal leakage. By synchronizing the entangling parameters, leakage channel, and the weakly frequency-dependent inter-qubit coupling, one can achieve high fidelity gates close to the speed limit set by the interaction strength. We demonstrate our approach with iSWAP-like and CPHASE gates in a frequency-tunable, fixed-coupling architecture, and find Pauli error rates down to $4.3(2)\cdot 10^{-3}$, corresponding to average gate fidelities up to $0.9966(2)$, for gate times as short as 1.2 times the speed limit. The presented approach can be useful for multibody operations as well.

We use frequency-tunable transmon qubits that are coupled through a fixed capacitance [Fig. 1(a)]. When tuning their transition frequencies near resonance populations can exchange, and the nonlinearity gives rise to a conditional phase accumulation as described (up to single-qubit phases) by the photon-conserving unitary

$$U = \begin{pmatrix} 1 & 0 & 0 & 0 \\ 0 & \cos\theta & -i\sin\theta & 0 \\ 0 & -i\sin\theta & \cos\theta & 0 \\ 0 & 0 & 0 & e^{-i\phi} \end{pmatrix} \quad (1)$$

with $\theta$ the swap angle and $\phi$ the conditional phase. However, when implementing such a gate, computational levels are unavoidably swept past or brought near resonance with higher levels, leading to leakage out of the computational subspace [Fig. 1(b)]. A complete description therefore involves the full one- and two-excitation subspace; the computational states $|01\rangle, |10\rangle, |11\rangle$ as well as the non-computational states $|02\rangle$ and $|20\rangle$.

The key notion behind our approach is minimizing the gate error by synchronizing the minima of the leakage and residual swap population, as suggested in Ref. [13]. For an ideal rectangular pulse of duration $t_h$, qubits are instantly placed on resonance and the states $|10\rangle$ and $|01\rangle$ will undergo Rabi oscillations with $P_{01}(t_h) = \sin^2(gt_h)$, and a complete population swap occurs for $t_h = \pi/(2g)$. Deviation from a full swap is given by $\epsilon_{\text{swap}} = 1 - P_{01}$. However, at the same time, $|11\rangle$ will interact with $|20\rangle$ and $|02\rangle$. The states $|20\rangle$ and $|02\rangle$ are in resonance and the three-level, two-excitation subspace is split into the "bright" subspace spanned by the states $|11\rangle$ and $|\psi_b\rangle = (|20\rangle + |02\rangle)/\sqrt{2}$, and the "dark state" $|\psi_d\rangle = (|20\rangle - |02\rangle)/\sqrt{2}$. The latter is decoupled from the other two states and remains unpopulated. The system undergoes off-resonant Rabi oscillations between the states $|11\rangle$ and $|\psi_b\rangle$, which are detuned from each other by the qubit nonlinearity $\eta$. In this ideal picture, the leakage error is simply given by the probability to occupy the non-computational state $|\psi_b\rangle$ at the end of the gate exe-cution,

$$\epsilon_{\text{leak}} = |\psi_b(t_h)|^2 = \frac{16g^2}{\eta^2 + 16g^2} \sin^2\left(\frac{1}{2}\sqrt{\eta^2 + 16g^2}\, t_h\right), \quad (2)$$

which is nulled for $\sqrt{\eta^2 + 16g^2}\, t_h = 2\pi n$, with $n$ an integer. Aligning the Rabi oscillations in the swap and leakage channels is the key behind the synchronization protocol.

To execute the iSWAP gate, one therefore needs to align a zero of the function $\epsilon_{\text{leak}}(t_h)$ with the maximum of $P_{01}(t_h)$. This implies that the inter-qubit coupling $g$ and the qubit nonlinearity $\eta$ must satisfy the relation

$$g = \frac{\eta}{4\sqrt{n^2 - 1}}, \quad (3)$$

and the synchronization is possible only for discrete values of $g$: the gate has a "spectrum". At first glance, the problem seems insurmountable with a fixed-coupling architecture. Furthermore, it is not clear if the above will hold for realistic pulses that are unavoidably rounded and broadened in experimental systems. In particular, a finite ramp speed could lead to an imperfect Landau-Zener transition through the $|11\rangle$ - $|20\rangle$ avoided level crossing, causing additional leakage not captured by Eq. (2).

The key to performing the synchronization lies in the architecture itself. We explore this experimentally using a pair of transmon qubits having maximum $f_{10}$ frequen-cies of 6.28 and 6.16 GHz and nearly constant nonlinearities $\eta/2\pi$ of 223 and 240 MHz. The coupling capacitor has a capacitance of $\sim 0.45$ fF, giving a coupling strength of $g/2\pi = 16.2$ MHz at $f_{10} = 6$ GHz, and is frequency-dependent following $g = \omega C_c/2C$.

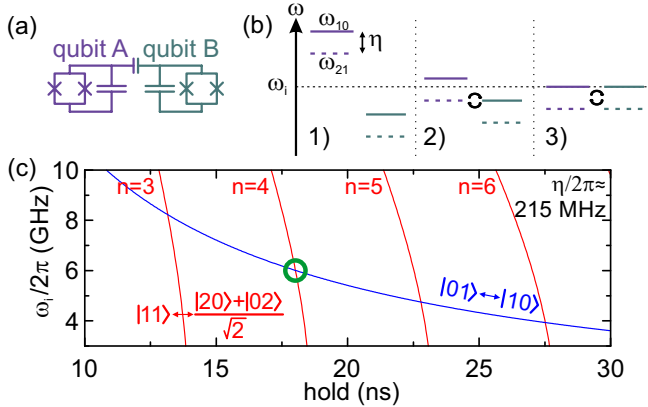

Figure 1. (a) Circuit diagram of two transmon qubits coupled through a small capacitance, giving rise to a frequency-dependent coupling strength. (b) Schematic representation of the iSWAP gate. 1) Qubits at rest (idle), separated by a large detuning. 2) As qubits are rapidly detuned to the interaction frequency, qubit B's $\omega_{10}$ transition sweeps by A's $\omega_{21}$ transition, potentially incurring leakage from $|11\rangle$ to $|20\rangle$. 3) Qubits at the interaction frequency undergo an iSWAP operation; both qubits' $|2\rangle$-states hybridize. Resonant Rabi oscillations occur between $|10\rangle$ and $|01\rangle$ while weaker off-resonant Rabi oscillations occur between $|11\rangle$ and $|\psi_b\rangle$. (c) The frequency dependence of the coupling $g$ makes the synchronization of the exchange between $|01\rangle$ and $|10\rangle$ and preservation of the occupation of $|11\rangle$ occur at specific frequencies.

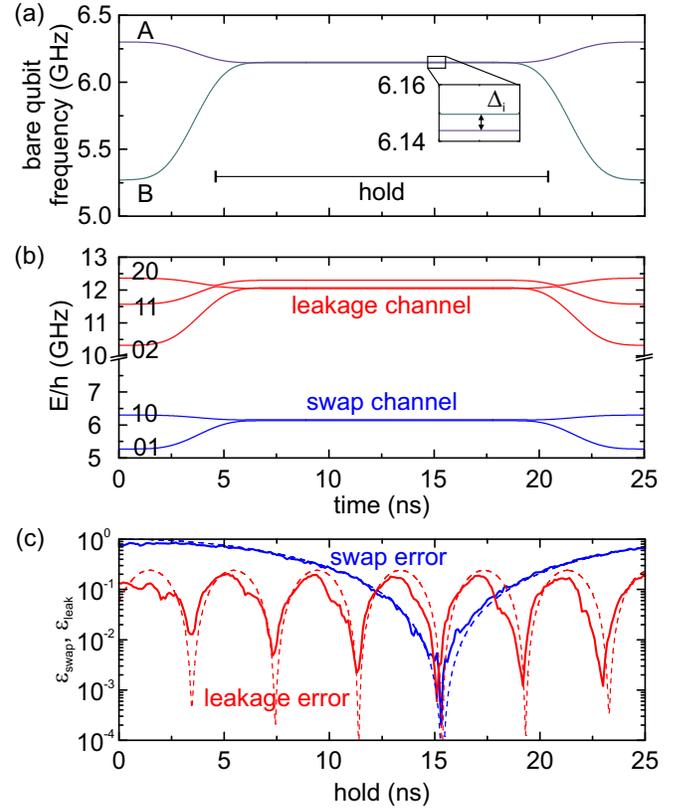

Figure 2. Achieving a swap and leakage error rate of below $10^{-3}$. (a) The pulse sequence consists of a rounded trapezoid that is applied on both qubit frequency control lines simultaneously to bring them to the interaction frequency. The inset highlights the small overshoot. Hold time is 15.2 ns. (b) During the hold period, swap (blue) and leakage channels (red) are realized. $|01\rangle$ and $|10\rangle$ are resonant, and $|11\rangle$ and $|\psi_b\rangle$ are near resonance. (c) By tuning the hold duration one can achieve synchronization of minimal swap and leakage error. Data not corrected for measurement visibility.

It is this frequency dependence that gives rise to spectra of synchronized swaps between $|01\rangle$ and $|10\rangle$ and swaps and back between $|11\rangle$ and $|\psi_b\rangle$ as a function of interaction frequency and hold time, shown in Fig. 1(c). These lines cross at several points that lie within the typical range of operation for superconducting qubits (4-7 GHz), indicating full synchronization between the swap and leakage channels.

Here, we focus on realizing this synchronization for $n = 4$, whose hold duration is shortest within the accessible range [green circle in Fig. 1(c)]. The control pulses are shown in Fig. 2(a). We simultaneously apply rounded trapezoidal flux pulses on the frequency control lines to steer the qubits towards the interaction frequency. We include a small overshoot $\Delta_i \sim 5$ MHz (inset), which acts as a fine-tuning parameter to account for nonidealities of the pulse, as well as the dressed eigenbasis due to small residual coupling before and after the duration of the gate. The overshoot comes from a complete swap requiring a rotation of the Bloch vector associated with the $\{|01\rangle, |10\rangle\}$ subspace by angle $\pi$. This Rabi oscillation is analogous to the Larmor precession of the spin in a magnetic field. When the qubits are idle there is a large $z$-component of the magnetic field due to the frequency detuning, and a small $x$-component due to $g$. As a result, the initial Bloch vector $\mathbf{m}(0)$ is slightly tilted towards the $x$-axis. At perfect resonance, the detuning is zero and the effective magnetic field is pointed along the $x$-axis, i.e. not perpendicular to $\mathbf{m}(0)$. Hence the Bloch vector will precess around a cone and will never point in the $-\mathbf{m}(0)$ direction, making a complete swap impossible. To enable a complete swap one needs to add a small $z$-field (overshoot).

The energy level diagram during the gate is plotted in Fig. 2(b), showing the swap channel, comprised of the hybridization of $|01\rangle$ and $|10\rangle$, and leakage channel, comprised of $|11\rangle$, $|02\rangle$, and $|20\rangle$.

We visualize the synchronization by varying the hold time and measuring the probabilities in the swap and leakage channels [Fig. 2(c)]. We initialize the qubits in either $|01\rangle$ or $|11\rangle$ and measure the unwanted probabilities $\epsilon_{\text{swap}}$ and $\epsilon_{\text{leak}}$, respectively. The leakage error (solid red line) shows dips every 4 ns, reaching a minimum of $6 \cdot 10^{-4}$ at 15.1 ns. The swap error (solid blue line) shows a single minimum of $2 \cdot 10^{-4}$ at 15.3 ns, overlapping with the 4th minimum of the leakage error. Data not corrected for measurement error. The dashed curves are theory predictions. We note that the minima of the leakage channel dip down to different values – a consequence of the qubits having dissimilar nonlinearities.

The data in Fig. 2(c) show that synchronization of leakage and swap error is achievable in an experimental setting, where control pulses are unavoidably rounded due to filtering and other distortions. As predicted in Fig. 1(c), synchronization occurs at $n = 4$. We find a small conditional phase of $\phi = 0.3 - 0.4$ rad that accumulates when qubit frequencies are steered towards and away from each other (not shown). Methods to control this phase will be published separately. Based on the measured probabilities of the swap and leakage error channels, this gate shows the potential to achieve an error rate below $10^{-3}$.

We explore the parameter range in Fig. 3, where the swap and leakage error are measured over interaction frequency, hold time, and overshoot. Model results are on the left, with experimental results on the right. Data vs interaction frequency and hold time show a relatively large frequency range over which synchronization can be achieved, see the white area in Fig. 3(a)-(b). This arises from the leakage error being only very weakly depen-dent on interaction frequency, evidenced by near-vertical blue lines (e.g. where leakage error is minimal), and a small dependence of the swap error on interaction frequency, consistent with Fig. 1c. The blue lines indicate regions of low leakage but high swap error, consistent with Fig. 2(c). Hold times are longer compared to Fig. 3 due to the use of different filtering. In Figs. 3(c)-(d) we see that the optimal value of the overshoot is chosen from a small range that is only weakly dependent on interaction frequency. We find a stronger tilt in the data in Fig. 3(d), compared to Fig. 3(c), which we attribute to minor distortions of the pulse waveform: the tilt is on the order of 10 MHz, while qubits are steered over a range of 1 GHz. Finally, Figs. 3(e)-(f) show that synchronization depends critically on both the overshoot frequency and hold time.

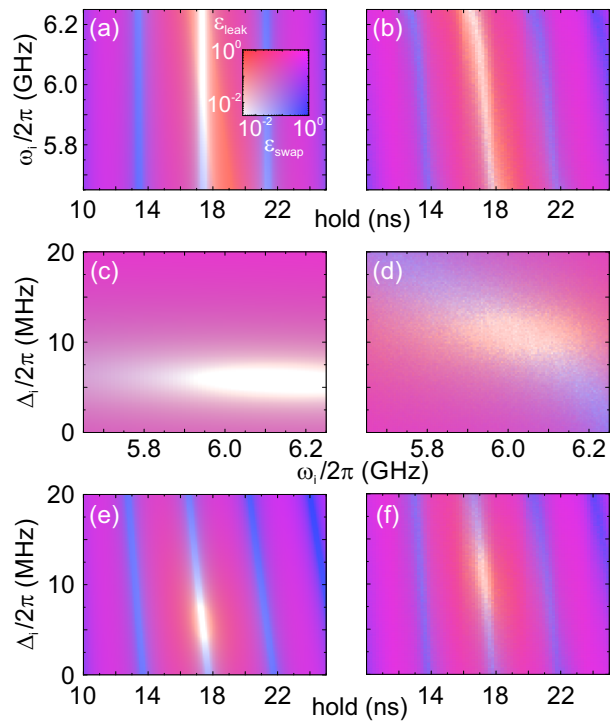

Figure 3. Mapping synchronization. Swap (blue) and leakage (red) error probabilities as a function of interaction frequency ($\omega_i$), overshoot frequency ($\Delta_i$), and hold times. Color legend is top left. Model results (a)(c)(e), and experimental results (b)(d)(f). Experimental results are not corrected for measurement error. Bright white areas indicate the parameter values for which the minima of leakage and swap error synchronize.

The robustness of both the hold time and overshoot frequency to the interaction frequency in Fig. 3 provides a clear map for tuning up and optimizing gate parameters. After choosing an interaction frequency, the optimal hold time and overshoot frequency can be clearly picked out from one-dimensional scans. We note that the inclusion of a small overshoot is critical to optimizing the synchronization, and is robust to the choice of interaction frequency and hold time.

Having found a good correspondence between theory and experiment in Fig. 3, we now turn to benchmarking the fidelity of the gate. We use cross-entropy benchmarking (XEB) [14], which can quantify the performance of a wider range of unitaries than Clifford-based randomized benchmarking. We follow the approach in Refs. [14, 15]. Here, we use two transmon qubits having maximum frequencies of 6.84 and 6.04 GHz, nonlinearities of 212 and 219 MHz, and $g/2\pi = 17.1$ MHz at the interaction frequency of 5.89 GHz. Qubit $T_1$ values lie around 20 $\mu$s. As the ideal interaction frequency lies slightly above the maximum frequency of the lower qubit, we bring up an iSWAP-like gate with a total duration of 18 ns, with incomplete swap angle $\theta = 1.42$ rad, see Fig. 4(a).

We can also implement a CPHASE gate [16], having a duration of 28 ns, that makes use of the interaction between $|11\rangle$ and $|20\rangle$ by choosing an overshoot close to the nonlinearity, intentionally swapping to $|20\rangle$ and then back [see Fig. 4(b)], underlining the generality of our approach.

XEB is a characterization tool based on sampling. Here, many cycles of gates are applied, with each cycle consisting of a layer of random single-qubit gates and the two-qubit gate, with a final round of random single-qubit gates appended at the end [inset Fig. 4(c)]. Single-qubit gates are chosen from the set of $\pi/2$ rotations around eight axes in the Bloch sphere representation: $\pm X$, $\pm Y$, and $\pm(X \pm Y)$. By comparing the measured state probabilities with the ideal ones we can define a sequence fidelity based on relative cross-entropy differences between probability distributions [14],

$$\alpha = \frac{H_{\text{inc, exp}} - H_{\text{meas, exp}}}{H_{\text{inc, exp}} - H_{\text{exp}}}. \quad (4)$$

Here, the cross entropy between two probability distributions $\{p_i\}$ and $\{q_i\}$ is $H = -\sum_i p_i \log(q_i)$, and $H_{\text{inc, exp}}$ is the cross-entropy between the incoherent (uniform) and expected (ideal) distributions, $H_{\text{meas, exp}}$ is the cross-entropy between the measured and expected distributions, and $H_{\text{exp}}$ is the self-entropy of the expected distribution. We sample over 100 different random circuits consisting of 500 cycles each. The decay in $\alpha$ with cycles reflects the accumulation of gate error, as in randomized benchmarking.

The sequence fidelity as a function of cycle number $m$ is shown in Fig. 4(c) for both gates, decaying according to $\alpha = Ap^m + B$ (solid lines), where state preparation and measurement errors are absorbed in $A$ and $B$. We find a Pauli error per cycle of $6.5(3) \cdot 10^{-3}$ for the iSWAP-like gate and $8(3) \cdot 10^{-3}$ for the CPHASE gate, where this quantity contains contributions from two single-qubit gates in addition to the two-qubit gate. We quantify the phases $\theta$ and $\phi$ of the unitary [Eq. (1)] using unitary tomography [17] and fine-tuning using the cross-entropy data to optimize the fidelity, finding $\theta = 1.42$ rad,

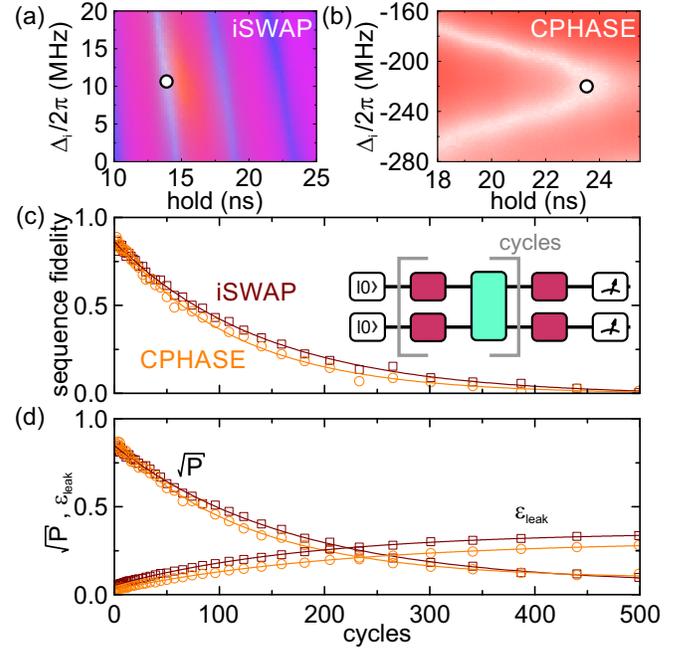

Figure 4. Gate benchmarking. (a) Swap and leakage error for iSWAP-like and (b) leakage error for CPHASE versus $\Delta_i$ and hold time. The white circle denotes parameters used for (c). Color legend from Fig. 3. (c) XEB cycle fidelity versus cycles using 100 random circuits. Inset: The qubits start in the ground state, XEB cycles are applied from the specific circuit, with each cycle comprised of single-qubit gates (red), and a two-qubit gate (green), appended by a final round of single qubit gates and measurement. This is repeated 1200 times for statistics per circuit. (d) Decay of $\sqrt{P}$ (square root of purity) and accumulation of leakage during XEB. Single-qubit XEB is in Ref. [17]. We extract Pauli gate errors of $4.3(2) \cdot 10^{-3}$ for the iSWAP-like and $5.8(2) \cdot 10^{-3}$ for the CPHASE gate, with contributions from control and decoherence of $1.8(4) \cdot 10^{-3}$ and $4.1(4) \cdot 10^{-3}$, respectively.

$\phi = 0.45$ rad for iSWAP-like and $\theta = 0.01$ rad, $\phi = 3.28$ rad for the CPHASE gate.

In order to separate out contributions to gate error, we have performed additional experiments, measuring the leakage and the purity $P$ [18], as well as performing single-qubit XEB. In Fig. 4(d) we plot the square root of the purity [19], which can be understood as the generalized Bloch vector length. It is therefore a measure of incoherent error, and is insensitive to coherent error. We find cycle purity errors of $5.5(1) \cdot 10^{-3}$ and $7.4(2) \cdot 10^{-3}$ for the iSWAP-like and CPHASE gate. In Fig. 4(d) we also plot the sum of all higher-level state populations accumulated during XEB in Fig. 4(c). Fitting to $\epsilon_{\text{leak}} = (p_0 - p_\infty)e^{-\Gamma m} + p_\infty$, with $\Gamma = \gamma_{\text{up}} + \gamma_{\text{down}}$ the sum of leakage and decay rates, respectively, $p_0$ the initial population, and $p_\infty = \gamma_{\text{up}}/\Gamma$ [8], we find leakage rates, expressed as Pauli errors of $2.43(4) \cdot 10^{-3}$ and $1.75(4) \cdot 10^{-3}$ for the iSWAP-like and CPHASE gate. The leaked population accumulates incoherently due to the interleaved random single-qubit gates, and therefore manifests as an incoherent error contribution [8], in both the purity and XEB benchmarking error results. Error is given by $r_{\mathrm{XEB}} - r_{\mathrm{purity}}$ for coherent error, and by $r_{\mathrm{purity}} - r_{\mathrm{leak}}$ for decoherence, with $r$ denoting the error extracted from the specific measurement.

The data in Fig. 4 indicate that the cycle errors for both gates are dominated by decoherence and leakage, that the leakage rate is around $2 \cdot 10^{-3}$ per cycle, and that control errors are around $1 \cdot 10^{-3}$ per cycle. We attribute the leakage per cycle being higher than observed for a single gate application in Fig. 2 to frequency control pulse distortions in long sequences. Separating out the single-qubit gate contribution we find Pauli errors of $4.3(2) \cdot 10^{-3}$ for the iSWAP-like and $5.8(2) \cdot 10^{-3}$, see [17] for the single-qubit data and gate error budget. These Pauli errors correspond to average gate fidelities of 0.9966(2) for the iSWAP-like and 0.9954(2) for the CPHASE gate. We find that these values are stable for over an hour [17]. These values compare well with those previously established for superconducting qubits: for CZ, average gate fidelities in the range to 0.990 - 0.994 having durations around 40 ns were reported [20, 21], and iSWAP gates in 180 ns have been shown with fidelities up to 0.98 [22]. Separate methods can be used for entangling gates in architectures with tunable coupling [23].

The demonstration of high fidelity iSWAP-like and CPHASE gates shows the viability of synchronization protocols to construct fast diabatic gates that have low leakage in a frequency-tunable qubit architecture. Using the native frequency dependencies in the system, this approach can be extended to simultaneously synchronizing multiple entanglement and leakage channels, enabling the construction of high-fidelity multi-qubit gates.

---

* These authors contributed equally to this work.
[1] A. Córcoles, E. Magesan, S. J. Srinivasan, A. W. Cross, M. Steffen, J. M. Gambetta, and J. M. Chow, Demonstration of a quantum error detection code using a square lattice of four superconducting qubits, Nature Communications **6** (2015).
[2] Y. Salathé, M. Mondal, M. Oppliger, J. Heinsoo, P. Kurpiers, A. Potočnik, A. Mezzacapo, U. Las Heras, L. Lamata, E. Solano, S. Filipp, and A. Wallraff, Digital quantum simulation of spin models with circuit quantum electrodynamics, Phys. Rev. X **5**, 021027 (2015).
[3] D. Ristè, S. Poletto, M.-Z. Huang, A. Bruno, V. Vesterinen, O.-P. Saira, and L. DiCarlo, Detecting bit-flip errors in a logical qubit using stabilizer measurements, Nature Communications **6** (2015).
[4] M. D. Reed, L. DiCarlo, S. E. Nigg, L. Sun, L. Frunzio, S. M. Girvin, and R. J. Schoelkopf, Realization of three-qubit quantum error correction with superconducting circuits, Nature **482**, 382 (2012).
[5] J. Kelly, R. Barends, A. G. Fowler, A. Megrant, E. Jeffrey, T. C. White, D. Sank, J. Y. Mutus, B. Campbell, Y. Chen, Z. Chen, B. Chiaro, A. Dunsworth, I. C. Hoi, C. Neill, P. J. J. O'Malley, C. Quintana, P. Roushan, A. Vainsencher, J. Wenner, A. N. Cleland, and J. M. Martinis, State preservation by repetitive error detection in a superconducting quantum circuit, Nature **519**, 66 (2015).
[6] R. Barends, A. Shabani, L. Lamata, J. Kelly, A. Mezzacapo, U. L. Heras, R. Babbush, A. G. Fowler, B. Campbell, Y. Chen, Z. Chen, B. Chiaro, A. Dunsworth, E. Jeffrey, E. Lucero, A. Megrant, J. Y. Mutus, M. Neeley, C. Neill, P. J. J. O'Malley, C. Quintana, P. Roushan, D. Sank, A. Vainsencher, J. Wenner, T. C. White, E. Solano, H. Neven, and J. M. Martinis, Digitized adiabatic quantum computing with a superconducting circuit, Nature **534**, 222 (2016).
[7] J. Koch, T. M. Yu, J. Gambetta, A. A. Houck, D. I. Schuster, J. Majer, A. Blais, M. H. Devoret, S. M. Girvin, and R. J. Schoelkopf, Charge-insensitive qubit design derived from the cooper pair box, Phys. Rev. A **76**, 042319 (2007).
[8] Z. Chen, J. Kelly, C. Quintana, R. Barends, B. Campbell, Y. Chen, B. Chiaro, A. Dunsworth, A. G. Fowler, E. Lucero, E. Jeffrey, A. Megrant, J. Mutus, M. Neeley, C. Neill, P. J. J. O'Malley, P. Roushan, D. Sank, A. Vainsencher, J. Wenner, T. C. White, A. N. Korotkov, and J. M. Martinis, Measuring and suppressing quantum state leakage in a superconducting qubit, Phys. Rev. Lett. **116**, 020501 (2016).
[9] A. G. Fowler, Coping with qubit leakage in topological codes, Phys. Rev. A **88**, 042308 (2013).
[10] J. M. Martinis and M. R. Geller, Fast adiabatic qubit gates using only $\sigma_z$ control, Phys. Rev. A **90**, 022307 (2014).
[11] R. Versluis, S. Poletto, N. Khammassi, B. Tarasinski, N. Haider, D. J. Michalak, A. Bruno, K. Bertels, and L. DiCarlo, Scalable quantum circuit and control for a superconducting surface code, Phys. Rev. Applied **8**, 034021 (2017).
[12] M. Ware, B. R. Johnson, J. M. Gambetta, T. A. Ohki, J. M. Chow, and B. L. T. Plourde, Cross-resonance interactions between superconducting qubits with variable detuning, arXiv preprint arXiv:1905.11480 (2019).
[13] F. W. Strauch, P. R. Johnson, A. J. Dragt, C. J. Lobb, J. R. Anderson, and F. C. Wellstood, Quantum logic gates for coupled superconducting phase qubits, Phys. Rev. Lett. **91**, 167005 (2003).
[14] C. Neill, P. Roushan, K. Kechedzhi, S. Boixo, S. V. Isakov, V. Smelyanskiy, A. Megrant, B. Chiaro, A. Dunsworth, K. Arya, *et al.*, A blueprint for demonstrating quantum supremacy with superconducting qubits, Science **360**, 195 (2018).
[15] S. Boixo, S. V. Isakov, V. N. Smelyanskiy, R. Babbush, N. Ding, Z. Jiang, M. J. Bremner, J. M. Martinis, and H. Neven, Characterizing quantum supremacy in near-term devices, Nature Physics **14**, 595 (2018).
[16] L. DiCarlo, M. D. Reed, L. Sun, B. R. Johnson, J. M. Chow, J. M. Gambetta, L. Frunzio, S. M. Girvin, M. H. Devoret, and R. J. Schoelkopf, Preparation and measurement of three-qubit entanglement in a superconducting circuit, Nature **467**, 574 (2010).
[17] See Supplemental Material.
[18] J. Wallman, C. Granade, R. Harper, and S. T. Flammia, Estimating the coherence of noise, New Journal of Physics **17**, 113020 (2015).

[19] G. Feng, J. J. Wallman, B. Buonacorsi, F. H. Cho, D. K. Park, T. Xin, D. Lu, J. Baugh, and R. Laflamme, Estimating the coherence of noise in quantum control of a solid-state qubit, Phys. Rev. Lett. **117**, 260501 (2016).

[20] M. A. Rol, F. Battistel, F. K. Malinowski, C. C. Bultink, B. M. Tarasinski, R. Vollmer, N. Haider, N. Muthusubramanian, A. Bruno, B. M. Terhal, *et al.*, A fast, low-leakage, high-fidelity two-qubit gate for a programmable superconducting quantum computer, arXiv preprint arXiv:1903.02492 (2019).

[21] R. Barends, J. Kelly, A. Megrant, A. Veitia, D. Sank, E. Jeffrey, T. C. White, J. Mutus, A. G. Fowler, B. Campbell, *et al.*, Superconducting quantum circuits at the surface code threshold for fault tolerance, Nature **508**, 500 (2014).

[22] D. C. McKay, S. Filipp, A. Mezzacapo, E. Magesan, J. M. Chow, and J. M. Gambetta, Universal gate for fixed-frequency qubits via a tunable bus, Phys. Rev. Applied **6**, 064007 (2016).

[23] B. Foxen *et al.*, Demonstrating a continuous set of two-qubit gates for near-term quantum algorithms, unpublished.

# Supplementary information for "Diabatic gates for frequency-tunable superconducting qubits"

R. Barends,[1,*] C. M. Quintana,[1,*] A. G. Petukhov,[2,*] Yu Chen,[1] D. Kafri,[2] K. Kechedzhi,[2] R. Collins,[1] O. Naaman,[1] S. Boixo,[2] F. Arute,[1] K. Arya,[1] D. Buell,[1] B. Burkett,[1] Z. Chen,[1] B. Chiaro,[3] A. Dunsworth,[3] B. Foxen,[3] A. Fowler,[1] C. Gidney,[1] M. Giustina,[1] R. Graff,[1] T. Huang,[1] E. Jeffrey,[1] J. Kelly,[1] P. V. Klimov,[1] F. Kostritsa,[1] D. Landhuis,[1] E. Lucero,[1] M. McEwen,[3] A. Megrant,[1] X. Mi,[1] J. Mutus,[1] M. Neeley,[1] C. Neill,[1] E. Ostby,[2] P. Roushan,[1] D. Sank,[1] K. J. Satzinger,[1] A. Vainsencher,[1] T. White,[1] J. Yao,[1] P. Yeh,[1] A. Zalcman,[1] H. Neven,[2] V. N. Smelyanskiy,[2] and John M. Martinis[1,3]

[1]*Google, Santa Barbara, CA 93117, USA*
[2]*Google, Venice, CA 90291, USA*
[3]*Department of Physics, University of California, Santa Barbara, CA 93106, USA*

## SINGLE-QUBIT CROSS-ENTROPY BENCHMARKING AND THE TWO-QUBIT GATE ERROR BUDGET

To estimate the two-qubit gate fidelity, we have performed cross-entropy benchmarking with the single qubits A and B. The single-qubit benchmarking experiments were performed immediately after the two-qubit gate benchmarking. We use 100 random circuits, using the single-qubit gateset formed by the eight $\pi/2$ rotations along the following axes in the Bloch sphere representation: $\pm X$, $\pm Y$, and $\pm(X \pm Y)$. Each cycle consists of a single application of a gate from the circuit, ending with a final random single-qubit gate and measurement. The cross-entropy sequence fidelity and the square root of purity are shown in Fig. S1a-b. We find similar decays for both the iSWAP-like and CPHASE case. Applying the analysis in the main text, we find XEB cycle Pauli errors of r=$1.9(1) \cdot 10^{-3}$ for qubit A and $1.12(4) \cdot 10^{-3}$ for qubit B. This indicates that on average, the Pauli error per gate is $1.6 \cdot 10^{-3}$. This compares well with previously obtained Pauli errors of single-qubit gates for our devices at $1.6 \cdot 10^{-3}$ from Ref. [1]. The purity Pauli errors, $1.30(8) \cdot 10^{-3}$ for qubit A and $1.04(4) \cdot 10^{-3}$ for qubit B per cycle, indicate that gate errors are predominantly limited by decoherence.

Using the single-qubit cross-entropy data we can estimate the average fidelity of the iSWAP-like and CPHASE gates. We can convert the decay parameter $p$ to an average error $r$ and Pauli error $r_P$, with the Pauli error being dimension-independent

$$r = \frac{N-1}{N}(1-p) \quad \text{(S1)}$$

$$r_P = \frac{N+1}{N}r \quad \text{(S2)}$$

with $N = 2^n$ the dimensionality of the system having $n$ qubits. Using that the Pauli fidelity $1 - r_P$ of a cycle is given by the product of gate Pauli gate fidelities: $(1 - r_{\text{cycle,P}}) = (1 - r_{\text{qA,P}})(1 - r_{\text{qB,P}})(1 - r_{\text{2qgate,P}})$, we can estimate the gate error of the two-qubit gate. We find Pauli errors of $4.3(2) \cdot 10^{-3}$ for the iSWAP-like and $5.8(2) \cdot 10^{-3}$. These can be converted into average gate fidelities $F = 1 - r$ (as usually reported by our field, but are dimension-dependent) of $0.9966(2)$ and $0.9954(2)$ for the iSWAP-like and CPHASE gate, respectively.

Following the same procedure for the purity, we extract purity errors of $3.8(2) \cdot 10^{-3}$ for the iSWAP-like gate and $5.6(3) \cdot 10^{-3}$ for the CPHASE gate.

The purity measurements directly probe the contribution from decoherence, it is also sensitive to leakage as higher level states are populated incoherently for our er-ror rates [2]. The leakage rate as determined in our experiment is averaged over input states and can be shown to directly contribute to gate error. Using the above we arrive at the error budget in Table S1. Values for the decay parameter $p$ are listed in Table S2.

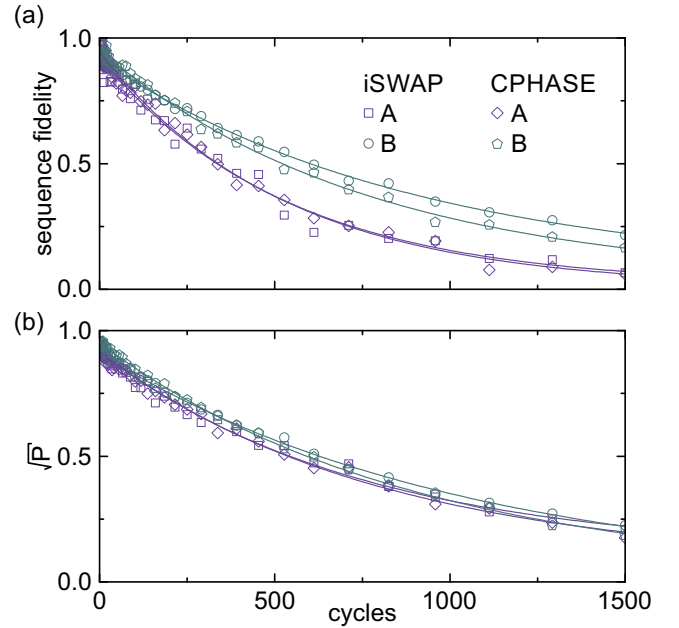

Figure S1. Single qubit XEB. (a) XEB fidelity decay as a function of the number of cycles, for qubit A and B, and for both the iSWAP-like and CPHASE experiment. For both gates we find the decay to be comparable and we extract a XEB cycle Pauli error of r=$1.9(1) \cdot 10^{-3}$ for qubit A and $1.12(4) \cdot 10^{-3}$ for qubit B. (b) The purity as a function of cycles. The purity decay is comparable for both qubits, giving a purity Pauli error of $1.30(8) \cdot 10^{-3}$ for qubit A and $1.04(4) \cdot 10^{-3}$. We find the leakage rate to be negligible (not shown).

Table S1. Gate error budget. Pauli errors in units of $10^{-3}$.

|  | iSWAP-like | CPHASE |
|---|---|---|
| decoherence error | 1.4(2) | 3.9(3) |
| leakage error | 2.43(4) | 1.75(4) |
| control error | 0.4(3) | 0.2(3) |
| total error | 4.3(2) | 5.8(3) |

Table S2. Decay parameter $p$ from fits to XEB decay and decay of the square root of purity $\sqrt{P}$ as shown in Fig. 4 in the main paper and Fig. S1.

|  | XEB | purity |
|---|---|---|
| iSWAP-like |  |  |
| two-qubit | 0.9931(2) | 0.9941(1) |
| qubit A | 0.9981(1) | 0.9987(1) |
| qubit B | 0.99889(5) | 0.99904(4) |
| CPHASE |  |  |
| two-qubit | 0.9913(3) | 0.9921(2) |
| qubit A | 0.9981(1) | 0.99874(9) |
| qubit B | 0.99870(7) | 0.99889(5) |

## TWO-QUBIT UNITARY

The generic photon-conserving two-qubit unitary for the iSWAP-like and CPHASE gate is given by

$$U = e^{-i(IZ-ZI)\delta_a/4} e^{-i(XX+YY)\theta/2} e^{-iZZ\phi/4}$$
$$\times e^{-i(IZ-ZI)\delta_b/4} e^{-i(IZ+ZI)\delta_+/4} \quad \text{(S3)}$$

specifying the swap angle $\theta$, conditional phase $\phi$, and separating out the non-commuting parts of the single-qubit phases, which naturally arise from the detuning operations that move qubits near and away from resonance. The phases $\delta_a$ and $\delta_b$ can be rewritten as $\delta_b = (\delta_c + \delta_d)/2$ and $\delta_a = (\delta_c - \delta_d)/2$. For iSWAP-like gates, the gate fidelity is only weakly dependent on the choice of $\delta_c$, because it is the phase of a diagonal matrix element that goes to zero for a full swap. Similarly, for CPHASE-like gates, the fidelity is only weakly dependent on $\delta_d$.

We determine the best-fit unitary model angles by minimizing the 2Q XEB cycle error with respect to the five angles using a simple Nelder-Mead optimization algorithm. The initial guess for this optimization is provided by a subset of two-qubit tomography measurements that provide good estimates of the angles. Full two-qubit process tomography is not required due to the known form of a generic photon-conserving two-qubit unitary.

## STABILITY OF THE TWO-QUBIT GATE PERFORMANCE

One of the key elements in the operation of gates is stability, as it allows for running algorithms for extended periods of time, as well as maintaining fidelity when tuning up multiple gates in series over large-scale systems. To quantify the stability, we have measured the cross-entropy benchmarking cycle error, purity error, and leakage rates for both the iSWAP-like and CPHASE gate for up to 2 hours, see Fig. S2. Gate parameters are constant, see Table S3.

Table S3. Gate parameters as defined in Eq. S3.

|  | $\theta$ | $\phi$ | $\delta_+$ | $\delta_c$ | $\delta_d$ |
|---|---|---|---|---|---|
| iSWAP-like | 1.42 | 0.48 | 2.02 | 4.34 | 4.39 |
| CPHASE | 0.01 | 3.29 | 0.39 | -0.13 | -3.96 |

We find that the cross-entropy cycle error closely follows the purity error, the purity error per cycle stays between $6 \cdot 10^{-3}$ and $8 \cdot 10^{-3}$, and that the leakage rate is virtually constant.

---

[1] R. Barends *et al.*, Nature **508**, pp. 500-503 (2014).
[2] Z. Chen *et al.*, Phys. Rev. Lett. **116**, 020501 (2016).

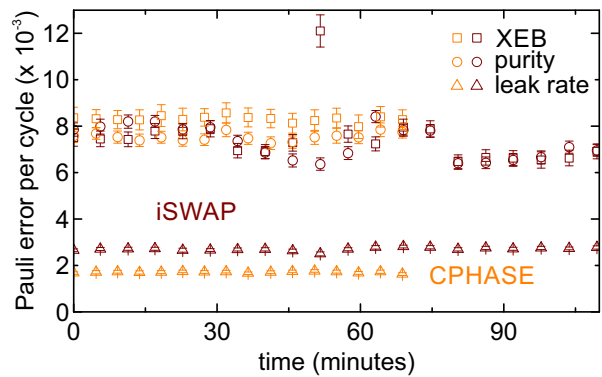

Figure S2. Cycle Pauli error stability. XEB cycle error, purity error and leakage rate versus time for both the iSWAP-like and CPHASE gates. The mean cycle control errors, computed by subtracting purity errors from cycle errors, are $0.3(3) \cdot 10^{-3}$ for the iSWAP-like and $0.78(4) \cdot 10^{-3}$ for the CPHASE gate.